 \documentclass[fleqn,12pt,twoside]{article}
 \usepackage{espcrc1}
 
 \usepackage{graphicx}
 \usepackage[figuresright]{rotating}
 
 \newcommand{\AmS}{{\protect\the\textfont2
   A\kern-.1667em\lower.5ex\hbox{M}\kern-.125emS}}
 
 \title{ Electromagnetic corrections to $\pi ^-$p scattering length from pionic hydrogen }
 
 \author{
  T.E.O.~Ericson \address{ CERN, CH-1211 Geneva 23, Switzerland, and 
 TSL, Box 533, S-75121 Uppsala, Sweden}, B.~Loiseau, \address{LPNHE, Universit\'{e} P. \& M. Curie,
 4 Place Jussieu, F-75252 Paris, France} and S.~Wycech \address{Soltan Institute for Nuclear Studies, PL-00681 Warszawa,
 Poland} }
        
\begin{document}
 
 \maketitle
 
 \begin{abstract}
We derive a closed, model space independent,  expression for the electromagnetic  correction factor $\delta $ to the scattering length  $a$
  extracted from  a hydrogenic atom with an extended charge to order $\alpha ^2$ and  $a^3$  in the limit of a short ranged hadronic  interaction. 
 
 \end{abstract}
 \vspace {4mm}

The strong interaction energy shifts $\epsilon _{1s}$ in pionic hydrogen and deuterium have recently been measured to a remarkable precision \cite{SCH01,HAUS98}.
 $$\epsilon _{1s}(\pi ^-p)=[-7.108\pm 0.013(stat)\pm0.034(syst)]~{\rm eV};~~\epsilon _{1s}(\pi ^-d)=[2.460\pm0.048]~{\rm eV}.$$
 
 It is well known\cite{DESER54,TRUMAN61}  that the strong interaction shift is intimately linked to the (complex)  scattering length  $a$
in the absence of the Coulomb field:
 \begin{equation}
 \epsilon _{1s}=4\pi /2m \ \phi ^2_B(0)\ a  \ (1+\delta ).
 \end{equation}
 Here $\phi_B(r)$ is the Bohr wave function and 
 $m$ the $\pi^-$p reduced mass. It is important  to understand the small correction  $\delta $  transparently and  reliably  to an accuracy matching the high experimental precision. This is desirable not only because the hadronic $\pi $N scattering lengths are key testing quantities for chiral physics, but also because they  are needed phenomenologically to a precision of about 1 \% \cite{ERICSON02}. 
 
 The present standard conversion of experimental data to a 
 scattering length uses the potential approach of Sigg et al.~\cite{SIGG96}, which describes the $\pi $N interaction in terms of coupled equations using physical pion masses and an isospin invariant non-diagonal potential matched to scattering lengths. This gives $\delta ({\rm Sigg})=-2.1\pm 0.5 \%$. This scattering length refers to a calculated  'isospin invariant' amplitude obtained  by setting the neutral pion mass equal to the charged one in the model.
 
 The approach has some weaknesses. First, the corresponding hadronic scattering amplitude does not reproduce properly the low energy s-wave scattering expansion as we have pointed out \cite{ERICSON02} and it is then not consistent with gauge invariance. Further, in the absence of a description for the dispersive correction from the radiative channel $\pi ^-p\rightarrow n \gamma $ they simulate it in an analogous way to that produced by charge exchange.  The correction to isospin invariance is model dependent. 
 
 Recently two calculations of the corrections inside Chiral Perturbation Theory (ChPT) have appeared with shockingly larger uncertainties. The next to leading order \cite{LYUBO00} gives $\delta ({\rm chiral }) =-4.3\pm 2.8~\%$ while  the next higher order \cite{GASSER02} give $\delta ({\rm chiral })=-7.2\pm 2.9~\%$. The main origin of the uncertainty is an anomalous photon-quark coupling. This large difference with Ref.\cite{SIGG96}   is mainly due to the definition of the scattering length in the chiral limit with the charged pion mass equal to the neutral one.  While this also gives an isospin invariant amplitude, it is now defined quite differently.
 
 In view of this situation we have carefully examined the extraction of the
 scattering length from the energy shift, its definition and corrections. The physics is largely governed by the low energy expansion of the $\pi $N amplitudes in the limit of isospin symmetry. In units of the charged pion mass $m_c$~\cite{HOEHLER}
 
 \begin{equation}
 [\tan\ (\delta_{0+})/q]_{\pi ^-p\rightarrow \pi ^-p}
 =(a^+_{0+}+a^-_{0+})+(b^+_{0+}+b^-_{0+})q^2+...
 \end{equation}
 
 \begin{equation}
 a^-_{0+}\simeq 0.09 m_c^{-1}~;~|a^+_{0+}|\simeq ({\rm few }~\%)\times a^-_{0+};~~~b^-_{0+}=0.013(6)m_c^{-3}~;~ b^+_{0+}=-0.044(7)m_c^{-3}. 
 \end{equation}

\noindent To elucidate the physics we use  both potential approaches and multiple scattering approaches together with analytical 
methods  {\it provided these reproduce the low energy expansion consistently}.
 Here our philosophy is close to that of Ref. \cite{LIPERTIA}, which takes the same attitude with respect to  the ChPT expansion, but we consider the phenomenological consistency with data  to be more important. One  notes that  a)  none of the previous approaches appears to be fully consistent in this perspective; furthermore none of them converts the energy shift $\epsilon _{1s} $ into a {\it physical} $\pi ^-p$ scattering length without the external Coulomb field;
 b)  the finite size shift is very small compared to the hadronic shift: $\epsilon _{fs}\sim -1.5\%{\rm ~ of~} \epsilon _{strong};$
 c)  this atomic problem in most aspects is highly non-relativistic and close to a long wave-length limit for the interaction.
 
 \noindent {\bf Corrections.}
 A number of the corrections are associated with the behavior of the unperturbed Bohr wave function at the origin:~$\phi _B({\vec r})=\phi _B(0)[1-\alpha 
 mr+..]$
 where the $r$-dependent term is  of the order of $10^{-2}$ at distances of the size of a hadron.
 
 \noindent {\bf Vacuum  polarisation correction.}
 The Uehling potential is of order $\alpha ^2$ but it gives a rather large contribution independent of the hadronic interaction such that $\epsilon _{vac}\sim 50\% {\rm ~of~ } -\epsilon _{1s}$.  It has a long range and can be seen as a modification of the Coulomb field with a slight change of the wave function $\delta \phi ^{vac}(0)$:
  \begin{equation}
 \delta _{vac}=2\delta  \phi ^{vac}(0)/\phi _B(0)=+0.48\%.
 \end{equation}
 This correction is very accurate. It has been  given  explicitly by Eiras and Soto \cite{EIR00}, but is implicitly included previously  \cite{SIGG96}. It is included in  the most recent ChPT effective field theory (EFT) approach \cite{GASSER02}, but not in earlier approaches.
 
 \noindent {\bf Leading corrections from the extended charge.}
 Previously Sigg et al.\cite{SIGG96} numerically investigated the shift in the presence of the combined charge distribution of the $\pi ^-$ and the proton with a charge radius $(\langle r^2_{\pi }\rangle +\langle r^2_p\rangle )^{1/2}\simeq 1.07 $ fm.  The poorly known range of the hadron interaction was varied. The EFT approaches do not clearly include  the accurately known electromagnetic form factors  and appear to replace its effects by a dimensional regularisation procedure and effective constants.

 Since the combined charge radius is larger than a typical hadronic interaction range it is natural to approach the problem from the limit of a zero range hadron interaction of scattering length $a$ as a leading approximation.
  The relation between scattering length and energy shift can then be obtained by matching the wave function to  the  scattering length condition at the origin. 
 We first observe that this problem has an exact solution
 for a charge placed on a sphere of radius $R$. The Coulomb potential inside this radius  is 
 constant, i.e., a square well, and consequently  the wave function is explicitly known.  By matching to the outside Coulomb wave function, the well known Whittaker function, the exact condition for the binding is obtained and we retain  the solution to order $\alpha ^2$. 
 
 We then consider the effect of the difference in the Coulomb potential between the soluble case and an actual charge distribution.  This interaction potential is perfectly regular and can be treated perturbatively. The result is the following closed expression for the scattering length correction, where $C$ is the Euler constant: 
 \begin{equation}
 \delta = -2m\alpha\ \langle r\rangle _{em}+2m\alpha\ a\ \left [2-C-\ln~(2\alpha) - \langle \ln~(mr)\rangle _{em} \right ].
 \end{equation}
 
 This relation ressembles superficially the Trueman formula, but the physics is quite different. The Trueman relation is expressed in terms of {\it hadronic } effective range and does not describe the finite size charge distribution which is the essential point here. There is no immediate relation between Trueman's effective range parameter   and the accurately known electromagnetic expectation values $\langle r\rangle _{em}$ and $\langle \ln~r\rangle _{em}$ above. The chiral approach is also different at this point since  it does not include the effect of the extended charge. 

 The terms in eq. (5) have a clear physical interpretation and are generated by the renormalization of the wave function at the origin. The first term appears since the wave function of the extended charge at the origin is 
 \begin{equation}
 \phi _{fs}(0)/\phi _B(0)~=~1-m\alpha \langle r\rangle _{em}+.. .
 \end{equation}
 This value for the wave function could have been used directly in the leading expression for the energy shift as a better first approximation than the wave function in the point Coulomb field.
 The second term proportional to the hadronic scattering length is due to the  effective incident field producing the hadron scattering and may be viewed as a cusp effect by the Coulomb field. We can show that the hadronic modification of the regular wave function at the origin is  exactly the one in the expression (5) above:
 \begin{equation}
 \psi _{reg}(0)/\phi _{fs}(0)~=~1+~2m\alpha ~ a~\left[2-C-\ln~(2\alpha ) -\langle \ln~(mr)\rangle _{em}\right]+.....
 \end{equation}
 
 Consequently to the order stated above the relation between the energy shift and the hadronic scattering length is fully understood in the limit of a zero range interaction.

\noindent {\bf Correction for the $q^2$ dependence of the $\pi $N amplitude}. 
 In the zero range limit the scattering amplitude  depends but trivially on the energy. The actual scattering  occurs in an attractive (Coulomb) potential of -2.5 MeV at the origin. The amplitude (2)  has  a momentum (energy)  dependence which  must included  consistently  respecting gauge invariance. The  corresponding expansion term $b$  is closely linked to the $\sigma $-term 
\cite {Ericson1987}. This term can be included \cite{Ericson1982} in an
energy representation using gauge invariance or as a momentum dependent
interaction to leading order.  Substituting $\omega \rightarrow \omega -eV_C(0)$ in the hadronic amplitude  (2) gives an additional term
\begin{equation}
a\ \delta _{q^2}=b~2m\alpha ~\langle 1/r\rangle _{em}.
\end{equation}
This term  expresses that the hadronic amplitude is evaluated at the local energy in the Coulomb field (but with the correct, nearly vanishing, physical phase space for scattering). With the parameters (3) for the $\pi ^ -p\rightarrow \pi ^-p$ scattering the correction is $-1.2\pm 0.4\%$, where the uncertainty is due to the experimental uncertainty of the phenomenological constants of Eq. (2). The extended charge is essential here, since this contribution otherwise will be divergent. This correction appears to correspond to the anomalous photon-quark coupling in the chiral EFT description, which represents a large source of uncertainty there.

The total correction in the determination of the scattering length with the external Coulomb field  switched off, is therefore $\delta _{total}=-1.1\pm0.4\%$. This is a small, well defined correction.  The scattering length contains isospin violating terms and 
is complex since both the physical $\pi ^0$n and the $\gamma $n channels are open. 

We can thus unambiguously convert the strong interaction
 energy shift into a  corresponding
scattering length. In the limit we consider, no model assumptions are needed and the resulting scattering length preserves the initial precision of the experiment. It is a separate problem to convert this physical quantity into into an isospin symmetric or chiral scattering length.

The $\pi ^-$p scattering length deduced here refers to scattering from a neutral system. It definition is closely analogous to the one in the  charge symmetric $\pi ^+$n channel, which is in principle directly measurable. The open decay channel and mass differences correspond exactly in the charge symmetry limit. This defines a natural starting point for the further discussion of its relation to isospin breaking and the chiral limit.

We are presently \cite{ERICSON03} investigating the sensitivity of our procedure to plausible assumptions of the  dependence of the physical hadronic amplitude on range etc. as well as the dispersive effects produced by the open decay channels. We recall that the scattering length $a$ with the external Coulomb field removed still contains internal Coulomb corrections such as mass differences and polarizability effects.

The  present method applies  with minor modifications to other pionic atoms, such as the $\pi ^+\pi  ^-$, the $\pi ^-K^+$ and $\pi ^-d$ systems.

 \end{document}